\documentclass[prb,twocolumn,aps,showpacs]{revtex4}
\usepackage{graphicx}%
\usepackage{dcolumn}
\usepackage{amsmath}
\usepackage{amssymb}
\usepackage{epsfig}

\begin{document}

\title{Bethe-Salpeter equations for the collective modes of the
$t-U-V-J$  model with d-wave pairing}
\author{Z. G. Koinov, P. Nash}\affiliation{Department of Physics and Astronomy,
University of Texas at San Antonio, San Antonio, TX 78249, USA}
\email{Zlatko.Koinov@utsa.edu} 

\begin{abstract}
The Bethe-Salpeter equations for the collective modes of a
$t$-$U$-$V$-$J$ model are used to analyze the resonance peak
observed at $\bf{Q}=(\pi,\pi)$ in neutron scattering experiments on
the cuprates. We assume that the resonance emerges due to the mixing
between the spin channel and 19 other channels. We have calculated
the energy of the lowest mode of the extended Hubbard model ($J=0$)
vs the on-site repulsive interaction $U$, as well as the $UJ$ lines
in the interaction parameter space which are consistent with the
ARPES data and reproduces the resonance peak at 40 meV in Bi2212
compound. We find that the resonance is predominantly a spin
exciton.
\end{abstract}

\maketitle \textbf{Introduction.} It is widely accepted that: (i)
the angle-resolved photoemission spectroscopy (ARPES) data produce
evidences for the opening of a d-wave pairing gap in cuprates
compounds described at low energies and temperatures by a BCS
theory, and  (ii) the basic pairing mechanism arises from the
antiferromagnetic exchange correlations, but the charge fluctuations
associated with double occupancy of a site also play an essential
role in doped systems. The simplest model that is consistent with
the last statements is the $t$-$U$-$V$-$J$ model. In the case of
d-pairing the gap function is $\Delta_\textbf{k}=\Delta
d_\textbf{k}/2$, where $\Delta$ is the maximum value of the energy
gap and $d_\textbf{k}=(\cos k_x-\cos k_y)$ (lattice constant $a=1$).
The BCS gap equation is
$1=\frac{V_\psi}{2}\int^\pi_{-\pi}\int^\pi_{-\pi}
\frac{d\textbf{k}}{(2\pi)^2} \frac{d^2_\textbf{k}}
{\sqrt{\overline{\varepsilon}^2_\textbf{k}+\Delta^2_\textbf{k}}}$,
where $V_\psi=2V+3J/2$,
$E(\textbf{k})=\sqrt{\overline{\varepsilon}^2_\textbf{k}+\Delta^2_\textbf{k}}$.
The mean-field electron energy $\overline{\varepsilon}_\textbf{k}$
has a tight-binding form
$\overline{\varepsilon}_\textbf{k}=t_1\left(\cos k_x+\cos
k_y\right)/2+t_2 \cos k_x\cos k_y +t_3(\cos 2k_x+\cos
2k_y)/2+t_4(\cos 2k_x\cos k_y+\cos 2k_y\cos k_x)/2+t_5\cos 2k_x\cos
2k_y -\mu$ obtained by fitting the ARPES data with a chemical
potential $\mu$ and hopping amplitudes $t_i$ for first to fifth
nearest neighbors on a square lattice.  $\Delta$, $t_1,...,t_5$ and
$\mu$ should all be thought of as an effective set of parameters,
while $V_\psi$ has to be determined by the gap equation. For Bi2212
compound, there are two possible sets of parameters with all
tight-binding basis functions involved (see Table 1 in Ref.
[\onlinecite{N}]). Assuming $\Delta=35$ meV, we obtain
$V^{(1)}_\psi=115.2$ meV with set 1, and
 $V^{(2)}_\psi=87.9$ meV with set 2.
Hao and Chubukov\cite{HC} have
 used another set of parameters (we shall call it H$\&$C) for Bi2212
compound with a doping concentration $x=0.12$: $t_1=-4t, t_2 =1.2t$,
$t=0.433$ eV, $\mu=-0.94t$, $\Delta=35$ meV and $V_\psi=0.6t$. The
parameters $U,V$ and $J$ should be adjusted in such a way that the
sharp collective mode which appears at wave vector
$\textbf{Q}_0=(\pi,\pi)$ in inelastic neutron-scattering resonance
(INSR) studies\cite{NSS} occurs at energy which corresponds to the
lowest collective mode of the corresponding Hamiltonian. In RPA the
resonance is determined by the pole of the spin correlation
function, which in the case of $J=0$ (the phase diagram at half
filling shows an "island" in U-V space where d-wave pairing
exists\cite{D}) is:
$\chi_s(\omega)=\chi^{0}_{00}(\textbf{Q}_0,\omega)/[1+U\chi^{0}_{00}(\textbf{Q}_0,\omega)]$,
where the bare spin correlation function\cite{BS,N} is
$\chi^{0}_{00}=I_{\widetilde{\gamma}\widetilde{\gamma}}$
($I_{\widetilde{\gamma}\widetilde{\gamma}}$ is defined later in the
text). Using the H$\&$C set of parameters\cite{HC} and a resonance
energy  of 40 meV, we calculate the RPA value of $U$ of about 1.16
eV. Sets 1 and 2 provide $U^{(1)}=0.533$ eV and $U^{(2)}=0.418$ eV,
respectively. The coupling of the spin channel with other channels
should change the RPA results for $U$. For example, we have two
$\pi$ channels\cite{Exc}  with bare $\pi$ susceptibilities
$\chi^{0}_{11}=I^{22}_{ll}$ and
$\chi^{0}_{22}=I^{22}_{\gamma\gamma}$, respectively. The
susceptibilities
 $I^2_{\gamma\widetilde{\gamma}},J^2_{l\widetilde{\gamma}},
J^{22}_{l\gamma}$ represent the mixing of the spin and two $\pi$
channels.  Thus, the coupling of the spin and two $\pi$ channels (a
three-channel response-function theory) leads in the generalized
random phase approximation (GRPA) to a set of three coupled
equations,\cite{HC} and the value of $U$ is reduced from 1.16 eV to
0.974 eV. When the extended spin channel is added to the previous
three channels, we have a set of four coupled equations (a
four-channel theory), and according to Ref. [\onlinecite{Plas}]
$U\approx 300$ meV is required in the case when $V_\psi$=0.260 eV
and $J=0$.

In what follows, the energy of the resonance is obtained from the
solution of 20 coupled Bethe-Salpeter (BS) equations for the
collective modes in GRPA, i.e.  the resonance emerges due to the
mixing between the spin channel and other 19 channels. In our
approach the INSR energy  solves
$det|\widehat{\chi}^{-1}-\widehat{V}|=0$, where the mean-field
response function $\widehat{\chi}$ and the interaction $\widehat{V}$
are $20\times 20$ matrices.  The secular determinant can be
rewritten as $det|\widehat{\chi}^{-1}-\widehat{V}|=det\left|
\begin{array}{cc}
A&B\\
B^T&C
\end{array}%
\right|=det|C|det|A-BC^{-1}B^T|$. In the case of the four-channel
response-function theory,\cite{Plas,Plas1} $A$ is a $4\times 4$
matrix while the mixing with the other 16 channels is represented by
a $4 \times 4$ matrix $BC^{-1}B^T$. We emphasize that none of the
previous theoretical interpretations of the INSR feature at
$\textbf{Q}_0$ have accounted properly for the mixing term
$BC^{-1}B^T$.

\textbf{t-U-V-J model.} The Hamiltonian of the $t$-$U$-$V$-$J$ model
consists of $t$ and $U$ terms representing the hopping of electrons
between sites of the lattice and their on-site repulsive
interaction, as well as the spin-independent attractive interaction
$V$ and the spin-dependent antiferromagnetic interaction $J$:
\begin{equation}\begin{split}
&H=-\sum_{i,j,\sigma}t_{ij}\psi^\dag_{i,\sigma}\psi_{j,\sigma}
-\mu\sum_{i,\sigma}\widehat{n}_{i,\sigma}+U\sum_i
\widehat{n}_{i,\uparrow} \widehat{n}_{i,\downarrow}\\&
-V\sum_{<i,j>\sigma\sigma'}\widehat{n}_{i,\sigma}\widehat{n}_{j,\sigma'}
+J\sum_{<i,j>}\overrightarrow{\textbf{S}}_i\textbf{.}\overrightarrow{\textbf{S}}_j.
\label{Hubb1}\end{split}\end{equation}  Here, the Fermi operator
$\psi^\dag_{i,\sigma}$ ($\psi_{i,\sigma}$) creates (destroys) a
fermion on the lattice site $i$ with spin projection
$\sigma=\uparrow,\downarrow$ along a specified direction and
$\widehat{n}_{i,\sigma}=\psi^\dag_{i,\sigma}\psi_{i,\sigma}$ is the
density operator on site $i$ with a position vector $\textbf{r}_i$.
The symbol $\sum_{<ij>}$ means sum over nearest-neighbor sites.
$t_{ij}$ is the single electron hopping integral. The
antiferromagnetic spin-dependent interaction $J
\sum_{<i,j>}\overrightarrow{\textbf{S}}_i\textbf{.}\overrightarrow{\textbf{S}}_j=J_1+J_2$
consists of two terms:
$J_1=\frac{J}{4}\sum_{<i,j>}[\widehat{n}_{i,\uparrow}\widehat{n}_{j,\uparrow}
+\widehat{n}_{i,\downarrow}\widehat{n}_{j,\downarrow}-
\widehat{n}_{i,\uparrow}\widehat{n}_{j,\downarrow}-
\widehat{n}_{i,\downarrow}\widehat{n}_{j,\uparrow}]$ and
$J_2=\frac{J}{2}\sum_{<i,j>}\left[\psi^\dag_{i,\uparrow}\psi_{i,\downarrow}
\psi^\dag_{j,\downarrow}\psi_{j,\uparrow}
+\psi^\dag_{i,\downarrow}\psi_{i,\uparrow}
\psi^\dag_{j,\uparrow}\psi_{j,\downarrow}\right]$.

It is useful to introduce four-component Nambu fermion fields
$\widehat{\overline{\psi}}
(y)=\left(\psi^\dag_\uparrow(y)\psi^\dag_\downarrow(y)\psi_\uparrow(y)\psi_\downarrow(y)
\right)$  and
$\widehat{\psi}(x)=\left(\psi^\dag_\uparrow(x)\psi^\dag_\downarrow(x)\psi_\uparrow(x)
\psi_\downarrow(x)\right)^T$, where $x$ and $y$ are composite
variables and the field operators obey anticommutation relations.
The "hat" symbol over any quantity $\widehat{O}$ means that this
quantity is a matrix.
\begin{figure}[tbp]
\includegraphics{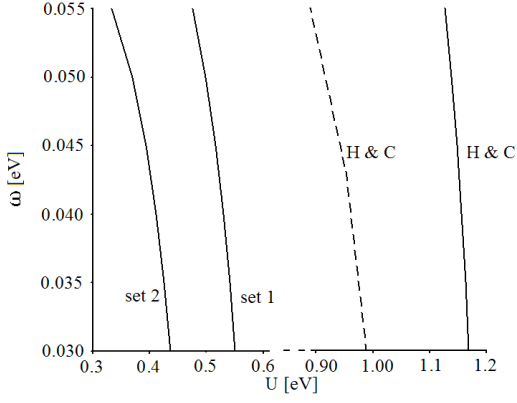} \label{Fig. 1}
\caption{The energy of the resonance obtained from the BS equations
when $J=0$. The curves are plotted using parameters  given in Table
1 in Ref. [\onlinecite{N}] (set 1 and set 2), and the Hao and
Chubukov parameters (curves H$\&$C).  The puncture curve represents
the three-channel energy (Fig. 4 in Ref. [\onlinecite{HC}]). }
\end{figure}

  The interaction
part of the extended Hubbard Hamiltonian is quartic in the Grassmann
fermion fields so the functional integrals cannot be evaluated
exactly.  However, we can transform the quartic terms to a quadratic
form by applying the Hubbard-Stratonovich transformation for the
electron operators:\cite{ZK1} $\int DA
e^{\left[\frac{1}{2}A_{\alpha}(z)D_{\alpha\beta}^{(0)-1}(z,z')A_{\beta}(z)+\widehat{\overline{\psi}}
(y)\widehat{\Gamma}^{(0)}_{\alpha}(y;x|z)\widehat{\psi}(x)A_{\alpha}(z)
\right]} =e^{-\frac{1}{2}\widehat{\overline{\psi}}
(y)\widehat{\Gamma}^{(0)}_{\alpha}(y;x|z)\widehat{\psi}(x)
D_{\alpha\beta}^{(0)}(z,z') \widehat{\overline{\psi}}
(y')\widehat{\Gamma}^{(0)}_{\beta}(y';x'|z')\widehat{\psi}(x')}$.
The last equation is used to define the $4\times 4$ matrices
$\widehat{D}_{\alpha\beta}^{(0)}$ and
$\widehat{\Gamma}^{(0)}_{\alpha}$ ($\alpha,\beta=1,2,3,4$). Their
Fourier transforms, written in terms of the Pauli  $\sigma_i$, Dirac
$\gamma^0$ and alpha\cite{Y,B} matrices, are as follows:
$\widehat{D}^{(0)}=\left(\begin{array}{cc}\widehat{D}_1&0\\0&\widehat{D}_2\end{array}%
\right)$, $\widehat{\Gamma}_{1,2}^{(0)}=(\gamma^0\pm\alpha_z)/2$ and
 $\widehat{\Gamma}_{3,4}^{(0)}=(\alpha_x\pm \imath\alpha_y)/2$, where
$\alpha_i=\left(\begin{array}{cc}\sigma_i&0\\0&\sigma_y\sigma_i\sigma_y
\end{array}%
\right)$,  $\widehat{D}_1=
\left(J(\textbf{k})-V(\textbf{k})\right)\sigma_0+\left(U-J(\textbf{k})-V(\textbf{k})\right)
\sigma_x$ and $\widehat{D}_2=2J(\textbf{k})\sigma_x$.   For a square
lattice and nearest-neighbor interactions
$V(\textbf{k})=4V(\cos(k_x)+\cos(k_y))$ and
$J(\textbf{k})=J(\cos(k_x)+\cos(k_y))$.  Now, we can establish a
one-to-one correspondence between the system under consideration and
a system
 which consists of a four-component boson field
$A_{\alpha}(z)$ interacting with fermion fields
$\widehat{\overline{\psi}} (y)$ and $\widehat{\psi}(x)$. The action
of the model system is $S= S^{(e)}_0+S^{(A)}_0+S^{(e-A)}$ where:
$S^{(e)}_0=\widehat{\overline{\psi}
}(y)\widehat{G}^{(0)-1}(y;x)\widehat{\psi} (x)$, $
S^{(A)}_0=\frac{1}{2}A_{\alpha}(z)D^{(0)-1}_{\alpha
\beta}(z,z')A_{\beta}(z')$ and $ S^{(e-A)}=\widehat{\overline{\psi}}
(y)\widehat{\Gamma}^{(0)}_{\alpha}(y,x\mid z)\widehat{\psi}
(x)A_{\alpha}(z)$. Here, we have used composite variables
$x,y,z=\{\textbf{r}_i,u\}$, where $\textbf{r}_{i}$ is a lattice site
vector, and  variable $u$ range from $0$ to $\beta=1/k_BT$ ($T$ and
$k_B$  are the temperature and the Boltzmann constant). We set
$\hbar=1$ and we use the summation-integration convention: that
repeated variables are summed up or integrated over.
\begin{figure}[tbp]
\includegraphics{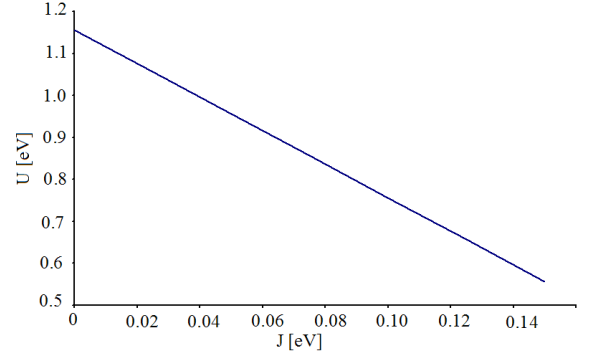} \label{Fig. 3}
\caption{Line in $U,J$ parameter space which reproduce the INSR
energy of 0.04 eV. Note that $V=V_\psi/2-3J/4$  where
$V_\psi=0.6t=259.8$ meV  is calculated from the gap equation by
using the set of parameters given in Ref.
[\onlinecite{HC}].}\end{figure}
\begin{figure}[tbp]
\includegraphics{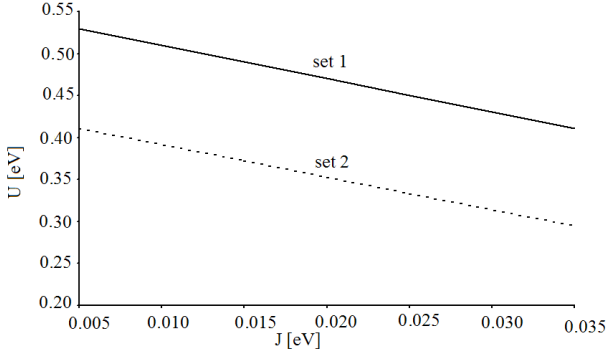} \label{Fig. 4}
\caption{Line in $U,J$ parameter space which reproduce the INSR
energy of 0.04 eV. The $V$ value is $V=V_\psi/2-3J/4$  where
$V^{(1)}_\psi=115.2$ meV and $V^{(2)}_\psi=87.9$ meV are calculated
by using two sets of parameters given in  Ref. [\onlinecite{N}].}
\end{figure}

Following the same steps as in  Refs. [\onlinecite{GFS,ZK}], we can
derive  a set of sixteen BS equations for the collective mode
$\omega(\textbf{Q})$ and BS amplitudes
$\Psi^\textbf{Q}_{n_1n_2}(\textbf{k})$ ($n_1,n_2 =1,2,3,4$). Their
matrix representation at zero temperature is :
\begin{equation}\begin{split}&
\widehat{\Psi}^\textbf{Q}(\textbf{k})=\frac{1}{N}\sum_{\textbf{q}}\int
\frac{d\Omega}{2\pi}\{-\widehat{D}^{(0)}_{\alpha\beta}(\textbf{k}-\textbf{q})\widehat{G}
(\textbf{k}+\textbf{Q}
;\Omega+\omega)\\&\widehat{\Gamma}^{(0)}_\alpha\widehat{\Psi}^\textbf{Q}(\textbf{q})
\widehat{\Gamma}^{(0)}_\beta
\widehat{G}(\textbf{k};\Omega)+\widehat{D}^{(0)}_{\alpha\beta}(\textbf{Q})\widehat{G}
(\textbf{k}+\textbf{Q}
;\Omega+\omega)\\&\widehat{\Gamma}^{(0)}_\alpha
\widehat{G}(\textbf{k};\Omega)Tr[
\widehat{\Gamma}^{(0)}_\beta\widehat{\Psi}^\textbf{Q}(\textbf{q})]\}
, \label{BS16}\end{split}\end{equation} where
$\widehat{G}(\textbf{k};\omega)$ is the BCS Green's
function.\cite{Y,B} The direct
$\widehat{D}^{(0)}_{\alpha\beta}(\textbf{k}-\textbf{q})
\widehat{\Gamma}^{(0)}_\alpha\widehat{\Gamma}^{(0)}_\beta$ and
exchange $\widehat{D}^{(0)}_{\alpha\beta}(\textbf{Q})
\widehat{\Gamma}^{(0)}_\alpha\widehat{\Gamma}^{(0)}_\beta$
interactions  mix all sixteen BS amplitudes. We can greatly simplify
Eqs. (\ref{BS16}) using the fact that in the RPA the
susceptibilities at $\textbf{Q}_0$ are convolutions of two
single-particle Green's functions $\widehat{G}$, and the equation
for the collective mode in the RPA is:
$\chi^{(0)-1}_1(\omega)\chi^{(0)-1}_2(\omega)-C_{12}(\omega)=0$,
where the susceptibilities $\chi^{(0)}_1$ and $\chi^{(0)}_2$
originate from ($U,J_1$) and $J_2$ interactions, respectively. The
term $C_{12}$ mixes the $J_1$ and $J_2$ interactions, but it is
proportional to convolutions which involve the anomalous Green's
functions $G_{13}$ and $G_{24}$. The two Green's functions appears
in the case of spin triplet pairing states where the order parameter
$\Delta_{\alpha\beta}(\textbf{k})$ is a $2\times 2$ matrix. For a
singlet superconductivity and d-wave pairing
$\Delta_{\alpha\beta}(\textbf{k})=i(\sigma_y)_{\alpha\beta}\Delta(\textbf{k})$,
$C_{12}(\omega)=0$, and the  equation for collective modes becomes
$\left[1+(U+4J)I_{\widetilde{\gamma}\widetilde{\gamma}}\right]
\left[1+4JI_{\widetilde{\gamma}\widetilde{\gamma}}\right]=0$, i.e.
$J_1$ and $J_2$ terms contribute separately to the collective modes.
Thus, we shall neglect all contributions due to the $J_2$ term in
Eqs. (\ref{BS16}). In this approximation we have a set of four
equations, which can be further simplified to a set of two equations
in the same manner as in Refs. [\onlinecite{GFS,ZK}]:
\begin{widetext}\begin{equation}
\begin{split}
&[\omega(\textbf{Q})-\varepsilon(\textbf{k},\textbf{Q})]G^{+}(\textbf{k},\textbf{Q})=
\frac{U}{2N}\sum_{\textbf{q}}
\left[\gamma_{\textbf{k},\textbf{Q}}\gamma_{\textbf{q},\textbf{Q}}+
l_{\textbf{k},\textbf{Q}}l_{\textbf{q},\textbf{Q}}\right]
G^{+}(\textbf{q},\textbf{Q})-\frac{U}{2N}\sum_{\textbf{q}}
\left[\gamma_{\textbf{k},\textbf{Q}}\gamma_{\textbf{q},\textbf{Q}}-
l_{\textbf{k},\textbf{Q}}l_{\textbf{q},\textbf{Q}}\right]
G^{-}(\textbf{q},\textbf{Q})\\&-\frac{1}{2N}\sum_{\textbf{q}}\left[V(\textbf{k}-\textbf{q})+
J(\textbf{k}-\textbf{q})\right]
\left[\gamma_{\textbf{k},\textbf{Q}}\gamma_{\textbf{q},\textbf{Q}}+
l_{\textbf{k},\textbf{Q}}l_{\textbf{q},\textbf{Q}}\right]
G^{+}(\textbf{q},\textbf{Q})\\&-\frac{1}{2N}\sum_{\textbf{q}}\left[V(\textbf{k}-\textbf{q})-
J(\textbf{k}-\textbf{q})\right] \left[
\widetilde{\gamma}_{\textbf{k},\textbf{Q}}
\widetilde{\gamma}_{\textbf{q},\textbf{Q}}+m_{\textbf{k},\textbf{Q}}m_{\textbf{q},\textbf{Q}}\right]
G^{+}(\textbf{q},\textbf{Q})\\&+\frac{1}{2N}\sum_{\textbf{q}}\left[V(\textbf{k}-\textbf{q})+
J(\textbf{k}-\textbf{q})\right]
\left[\gamma_{\textbf{k},\textbf{Q}}\gamma_{\textbf{q},\textbf{Q}}-
l_{\textbf{k},\textbf{Q}}l_{\textbf{q},\textbf{Q}}\right]
G^{-}(\textbf{q},\textbf{Q})\\&+\frac{1}{2N}\sum_{\textbf{q}}\left[V(\textbf{k}-\textbf{q})-
J(\textbf{k}-\textbf{q})\right] \left[
\widetilde{\gamma}_{\textbf{k},\textbf{Q}}
\widetilde{\gamma}_{\textbf{q},\textbf{Q}}-m_{\textbf{k},\textbf{Q}}m_{\textbf{q},\textbf{Q}}\right]
G^{-}(\textbf{q},\textbf{Q})\\&-\frac{U-2J(\textbf{Q})}{2N}\sum_{\textbf{q}}
\widetilde{\gamma}_{\textbf{k},\textbf{Q}}\widetilde{\gamma}_{\textbf{q},\textbf{Q}}
\left(G^{+}(\textbf{q},\textbf{Q})
-G^{-}(\textbf{q},\textbf{Q})\right)+\frac{U-2V(\textbf{Q})}{2N}\sum_{\textbf{q}}m_{\textbf{k},\textbf{Q}}m_{\textbf{q},\textbf{Q}}
\left[G^{+}(\textbf{q},\textbf{Q})+G^{-}(\textbf{q},\textbf{Q})\right],
\label{NewEq1}
\end{split}
\end{equation}
\begin{equation}
\begin{split}
&[\omega(\textbf{Q})+\varepsilon(\textbf{k},\textbf{Q})]G^{-}(\textbf{k},\textbf{Q})=
-\frac{U}{2N}\sum_{\textbf{q}}
\left[\gamma_{\textbf{k},\textbf{Q}}\gamma_{\textbf{q},\textbf{Q}}+
l_{\textbf{k},\textbf{Q}}l_{\textbf{q},\textbf{Q}}\right]
G^{-}(\textbf{q},\textbf{Q})+\frac{U}{2N}\sum_{\textbf{q}}
\left[\gamma_{\textbf{k},\textbf{Q}}\gamma_{\textbf{q},\textbf{Q}}-
l_{\textbf{k},\textbf{Q}}l_{\textbf{q},\textbf{Q}}\right]
G^{+}(\textbf{q},\textbf{Q})\\&+\frac{1}{2N}\sum_{\textbf{q}}\left[V(\textbf{k}-\textbf{q})+
J(\textbf{k}-\textbf{q})\right]
\left[\gamma_{\textbf{k},\textbf{Q}}\gamma_{\textbf{q},\textbf{Q}}+
l_{\textbf{k},\textbf{Q}}l_{\textbf{q},\textbf{Q}}\right]
G^{-}(\textbf{q},\textbf{Q})\\&+\frac{1}{2N}\sum_{\textbf{q}}\left[V(\textbf{k}-\textbf{q})-
J(\textbf{k}-\textbf{q})\right] \left[
\widetilde{\gamma}_{\textbf{k},\textbf{Q}}
\widetilde{\gamma}_{\textbf{q},\textbf{Q}}+m_{\textbf{k},\textbf{Q}}m_{\textbf{q},\textbf{Q}}\right]
G^{-}(\textbf{q},\textbf{Q})\\&-\frac{1}{2N}\sum_{\textbf{q}}\left[V(\textbf{k}-\textbf{q})+
J(\textbf{k}-\textbf{q})\right]
\left[\gamma_{\textbf{k},\textbf{Q}}\gamma_{\textbf{q},\textbf{Q}}-
l_{\textbf{k},\textbf{Q}}l_{\textbf{q},\textbf{Q}}\right]
G^{+}(\textbf{q},\textbf{Q})\\&-\frac{1}{2N}\sum_{\textbf{q}}\left[V(\textbf{k}-\textbf{q})-
J(\textbf{k}-\textbf{q})\right] \left[
\widetilde{\gamma}_{\textbf{k},\textbf{Q}}
\widetilde{\gamma}_{\textbf{q},\textbf{Q}}-m_{\textbf{k},\textbf{Q}}m_{\textbf{q},\textbf{Q}}\right]
G^{+}(\textbf{q},\textbf{Q})\\&
-\frac{U-2J(\textbf{Q})}{2N}\sum_{\textbf{q}}
\widetilde{\gamma}_{\textbf{k},\textbf{Q}}\widetilde{\gamma}_{\textbf{q},\textbf{Q}}
\left(G^{+}(\textbf{q},\textbf{Q})
-G^{-}(\textbf{q},\textbf{Q})\right)-\frac{U-2V(\textbf{Q})}{2N}\sum_{\textbf{q}}m_{\textbf{k},\textbf{Q}}m_{\textbf{q},\textbf{Q}}
\left[G^{+}(\textbf{q},\textbf{Q})+G^{-}(\textbf{q},\textbf{Q})\right].
\label{NewEq2}
\end{split}
\end{equation}Here
$\varepsilon(\textbf{k},\textbf{Q})=
E(\textbf{k}+\textbf{Q})+E(\textbf{k})$, and we use the same form
factors as in Ref.[\onlinecite{GFS}]:
$\gamma_{\textbf{k},\textbf{Q}}=u_{\textbf{k}}u_{\textbf{k}+\textbf{Q}}+v_{\textbf{k}}v_{\textbf{k}+\textbf{Q}},\quad
l_{\textbf{k},\textbf{Q}}=u_{\textbf{k}}u_{\textbf{k}+\textbf{Q}}-v_{\textbf{k}}v_{\textbf{k}+\textbf{Q}},
\quad
\widetilde{\gamma}_{\textbf{k},\textbf{Q}}=u_{\textbf{k}}v_{\textbf{k}+\textbf{Q}}-u_{\textbf{k}+\textbf{Q}}v_{\textbf{k}},
$ and $ m_{\textbf{k},\textbf{Q}}=
u_{\textbf{k}}v_{\textbf{k}+\textbf{Q}}+u_{\textbf{k}+\textbf{Q}}v_{\textbf{k}}
$ where $u^2_{\textbf{k}}=1-v^2_{\textbf{k}}=
\left[1+\overline{\varepsilon}(\textbf{k})/E(\textbf{k})\right]/2$.

It is worth mentioning that in the case of an extended Hubbard model
($J=0$), Eqs. (\ref{NewEq1}) and (\ref{NewEq2}) are the exact BS
equations in the GRPA. They are in accordance with the Goldstone
theorem which says that the gauge invariance is restored by the
existence of the Goldstone mode whose energy approaches zero at
$\textbf{Q}=0$. The last statement corresponds to the so-called
trivial solution of the BS equations:
$G^{+}(\textbf{k},\textbf{Q}=0)=-G^{-}(\textbf{k},\textbf{Q}=0)
=\Delta_\textbf{k}/2E(\textbf{k})$, and the gap equation\cite{R}
$\Delta_\textbf{k}=\frac{1}{N}\sum_\textbf{q}
[-U+V(\textbf{k}-\textbf{q})]\Delta_\textbf{q}/2E(\textbf{q})$ is
recovered  from our BS equations.

  The Fourier
transforms of  $V$ and $J$ interactions are separable, i.e.
$V(\textbf{k}-\textbf{q})=2V\widehat{\lambda}_\textbf{k}\widehat{\lambda}^T_\textbf{q}$
and
$J(\textbf{k}-\textbf{q})=J\widehat{\lambda}_\textbf{k}\widehat{\lambda}^T_\textbf{q}/2$,
and therefore, Eqs. (\ref{NewEq1}) and (\ref{NewEq2}) can be solved
analytically. Here
$\widehat{\lambda}_\textbf{k}=\left(s_\textbf{k},d_\textbf{k},ss_\textbf{k},sd_\textbf{k}\right)$
 is an $1\times 4$ matrix, and we have used the following notations:
$s_\textbf{k}=\cos(k_x)+\cos(k_y)$,
$d_\textbf{k}=\cos(k_x)-\cos(k_y)$,$ss_\textbf{k}=\sin(k_x)+\sin(k_y)$
and $cd_\textbf{k}=\sin(k_x)-\sin(k_y)$. Thus, we obtain a set of 20
coupled linear homogeneous equations for the dispersion of the
collective excitations. The existence of a non-trivial solution
requires that the secular determinant
$det\|\widehat{\chi}^{-1}-\widehat{V}\|$ is equal to zero, where the
bare mean-field-quasiparticle response function
$\widehat{\chi}=\left(
\begin{array}{cc}
P&Q\\
Q^T&R
\end{array}%
\right)$  and the interaction
$\widehat{V}=diag(U,U,-(U+4J),U+16V,-(2V+J/2),...,-(2V+J/2),-(2V-J/2),...,-(2V-J/2))$
are $20\times 20$ matrices. $P$ and $Q$ are $4\times 4$ and $4\times
16$ blocks, respectively, while $R$ is $16\times 16$ block (in what
follows $i,j=1,2,3,4$):\begin{equation} P=\left|
\begin{array}{cccc}
I_{\gamma,\gamma}&J_{\gamma,l}&I_{\gamma,\widetilde{\gamma}}&J_{\gamma,m}\\
J_{\gamma,l}&I_{l,l}&J_{l,\widetilde{\gamma}}&I_{l,m}\\
I_{\gamma,\widetilde{\gamma}}&J_{l,\widetilde{\gamma}}&
I_{\widetilde{\gamma},\widetilde{\gamma}}&
J_{\widetilde{\gamma},m}\\
J_{\gamma,m}&I_{l,m}&J_{\widetilde{\gamma},m}&I_{m,m}
\end{array}%
\right|,Q=\left|
\begin{array}{cccc}
I^i_{\gamma,\gamma}&J^i_{\gamma,l}&I^i_{\gamma,\widetilde{\gamma}}&J^i_{\gamma,m}\\
J^i_{\gamma,l} &I^i_{l,l}&J^i_{l,\widetilde{\gamma}}&I^i_{l,m}\\
I^i_{\gamma,\widetilde{\gamma}}& J^i_{l,\widetilde{\gamma}}&I^i_{\widetilde{\gamma},\widetilde{\gamma}}&J^i_{\widetilde{\gamma},m}\\
J^i_{\gamma,m}&I^i_{l,m} &J^i_{\widetilde{\gamma},m} &I^i_{m,m}
\end{array}%
\right|, R=\left|
\begin{array}{cccc}
I^{ij}_{\gamma, \gamma}&J^{ij}_{\gamma, l}& I^{ij}_{\gamma,
\widetilde{\gamma}}&J^{ij}_{\gamma,
 m}\\
J^{ij}_{\gamma, l} &I^{ij}_{l,l}&
J^{ij}_{l,\widetilde{\gamma}}&I^{ij}_{l,m}\\
 I^{ij}_{\gamma
\widetilde{\gamma}}& J^{ij}_{l,\widetilde{\gamma}}&
 I^{ij}_{\widetilde{\gamma},\widetilde{\gamma}}&
J^{ij}_{\widetilde{\gamma}, m}\\
 J^{ij}_{\gamma,
 m}&I^{ij}_{l,m} &J^{ij}_{\widetilde{\gamma}, m}
  &I^{ij}_{m,m}
\end{array}%
\right|.\nonumber\end{equation}  The quantities
$I_{a,b}=F_{a,b}(\varepsilon (\mathbf{k},\mathbf{Q}))$ and
$J_{a,b}=F_{a,b}(\omega)$, the $1\times 4$ matrices
$I^i_{a,b}=F^{i}_{a,b}(\varepsilon (\mathbf{k},\mathbf{Q}))$ and
$J^i_{a,b}=F^{i}_{a,b}(\omega)$, and the $4\times 4$ matrices
$I^{ij}_{a,b}=F^{ij}_{a,b}(\varepsilon (\mathbf{k},\mathbf{Q}))$ and
$J^{ij}_{a,b}=F^{ij}_{a,b}(\omega)$ are defined as follows (the quantities $a(\mathbf{k}%
,\mathbf{Q})$ and $b(\mathbf{k},\mathbf{Q})=l_{\mathbf{k},\mathbf{Q}},m_{%
\mathbf{k},\mathbf{Q}},\gamma _{\mathbf{k},\mathbf{Q}}$ or $\widetilde{%
\gamma }_{\mathbf{k},\mathbf{Q}}$):
$$
F_{a,b}(x)\equiv \frac{1}{N}\sum_\textbf{k}\frac{%
xa(\mathbf{k},\mathbf{Q})b(\mathbf{k},%
\mathbf{Q})}{\omega ^{2}-\varepsilon ^{2}(\mathbf{k},\mathbf{Q})},
F^i_{a,b}(x)\equiv \frac{1}{N}\sum_\textbf{k}\frac{%
xa(\mathbf{k},\mathbf{Q})b(\mathbf{k},%
\mathbf{Q})\widehat{\lambda}^i_\textbf{k}}{\omega ^{2}-\varepsilon
^{2}(\mathbf{k},\mathbf{Q})},F^{ij}_{a,b}(x)\equiv \frac{1}{N}\sum_\textbf{k}\frac{%
xa(\mathbf{k},\mathbf{Q})b(\mathbf{k},%
\mathbf{Q})}{\omega ^{2}-\varepsilon
^{2}(\mathbf{k},\mathbf{Q})}\left(\widehat{\lambda}^T_\textbf{k}
\widehat{\lambda}_\textbf{k}\right)_{ij}.$$ \end{widetext} The
elements of $P, Q$ and $R$ blocks are convolutions of conventional
two normal $GG$, two anomalous $FF$ Green's functions or $FG$ terms.
At the high-symmetry wave vector $\textbf{Q}_0$, $I_{a,b}^{i}$ and
$J_{a,b}^{i}$ with $i=3,4$ involve sine functions, and therefore,
all vanish. $I_{a,b}^2$ and $J_{a,b}^2$ also vanish because
$\varepsilon (\mathbf{k},\mathbf{Q}_0)$ is symmetric with respect to
exchange $k_x\leftrightarrow k_y$. Similarly, the non-diagonal
elements of $I_{a,b}^{ij}$ and $J_{a,b}^{ij}$ with $i\neq j$ all
vanish. Thus, blocks $P$ and $Q$, each has 10 different non-zero
elements, while  $R$ has 40 non-zero elements. In other words, the
$\omega$ dependence of $\widehat{\chi}$ (or $\widehat{\chi}^{-1}$)
comes from these 60 non-zero elements. It is worth mentioning that
within the four-channel theory\cite{Plas} the collective mode energy
has been calculated by using a $4\times 4$ symmetric matrix
$\widehat{\chi}$ which has only 6 non-zero elements at $\bf{Q}_0$:
$\widehat{\chi}_{11}= I_{\widetilde{\gamma}\widetilde{\gamma}},
\widehat{\chi}_{22}= I_{mm}^{11}, \widehat{\chi}_{33}=
I_{\gamma\gamma}^{22},\widehat{\chi}_{44}= I_{ll}^{22},
\widehat{\chi}_{12}= J^1_{m\widetilde{\gamma}}$ and
$\widehat{\chi}_{34}= J^{22}_{l\gamma}$ (the other 4 elements
$\widehat{\chi}_{13}=
I^2_{\gamma\widetilde{\gamma}},\widehat{\chi}_{14}=
J^2_{l\widetilde{\gamma}}, \widehat{\chi}_{23}= J^{12}_{m\gamma}$
and $\widehat{\chi}_{24}= I^{12}_{ml}$ vanish).

In Fig. 1 we present the results of our calculations of the lowest
collective mode of the extended Hubbard model ($J=0$) using
$49\times 49$ $\bf{k}$ points in the Brillouin zone and three
possible sets of parameters: sets 1 and 2 include  all tight-binding
basis functions (see Table 1 in Ref. [\onlinecite{N}]), while the
third set (H$\&$C) is used by Hao and Chubukov.\cite{HC} As can be
seen in Fig. 1, BS equations provide energies which are
significantly different from those obtained according to the
three-channel theory (see Fig. 4 in Ref.[\onlinecite{HC}]). In Fig.
2 and Fig. 3 we present the results of our calculations of the lines
in $U,J$ parameter space which reproduce the INSR energy of 40 meV
using all twenty channels. We see that the RPA spin correlation
function  and the BS equations in GRPA, both provide very similar
results for $U$ at point $J=0$. This indicates that the resonance
remains predominantly a spin exciton.

In summary, we have derived a set of four coupled BS equations for
the collective modes of the $t-U-V-J$ model including the $J_1$ part
of the antiferromagnetic interaction. These equations have been used
to analyze the resonance peak in Bi2212. It is interesting
 to note that  the trivial solution  of
 the BS equations (\ref{NewEq1}) and (\ref{NewEq2} leads to an equation similar to the  gap equation  but
 with $V_\psi=2V+J/2$ instead of $V_\psi=2V+3J/2$. The Goldstone mode, which is expected on physical grounds as the
symmetry is spontaneously broken by the condensate, does exist as a
trivial solution of the sixteen BS equations.

\end{document}